\documentclass[conference]{IEEEtran}

\usepackage{cite}

\ifCLASSINFOpdf
  \usepackage[pdftex]{graphicx}
\else
  \usepackage[dvips]{graphicx}
\fi
\usepackage{epstopdf}
\usepackage{graphicx}
\usepackage[ruled]{algorithm}

\usepackage[utf8]{inputenc}
\usepackage{color}
\usepackage{colortbl}
\usepackage[usenames,dvipsnames]{xcolor}
\usepackage{pgf,tikz}
\usetikzlibrary{arrows, automata}
\usepackage{multirow}
\usepackage{verbatim}
\usepackage{algorithmicx}
\usepackage{algpseudocode}

\newcommand{\LineComment}[1][\footnotesize ]{\Statex \hfill\(\triangleright\) #1}
\algnewcommand\algorithmicswitch{\textbf{switch}}
\algnewcommand\algorithmiccase{\textbf{case}}
\algnewcommand\algorithmicassert{\texttt{assert}}
\algnewcommand\Assert[1]{\State \algorithmicassert(#1)}%
\algdef{SE}[SWITCH]{Switch}{EndSwitch}[1]{\algorithmicswitch\ #1\ \algorithmicdo}{\algorithmicend\ \algorithmicswitch}%
\algdef{SE}[CASE]{Case}{EndCase}[1]{\algorithmiccase\ #1}{\algorithmicend\ \algorithmiccase}%
\algtext*{EndSwitch}%
\algtext*{EndCase}%

\input{mathematica.data}

\begin{document}

\title{Context-Aware Adaptive Framework\\for e-Health Monitoring}

\author{\IEEEauthorblockN{Haider Mshali}
\IEEEauthorblockA{University of Bordeaux -- LaBRI\\
Talence, France\\
haider-hasan.mshali@u-bordeaux.fr}
\and
\IEEEauthorblockN{Tayeb Lemlouma}
\IEEEauthorblockA{University of Rennes 1 -- IRISA\\
Lannion, France\\
tayeb.lemlouma@irisa.fr}
\and
\IEEEauthorblockN{Damien Magoni}
\IEEEauthorblockA{University of Bordeaux -- LaBRI\\
Talence, France\\
magoni@labri.fr}}

\maketitle

\begin{abstract}
For improving e-health services, we propose a context-aware framework to monitor the activities of daily living of dependent persons. We define a strategy for generating long-term realistic scenarios and a framework containing an adaptive monitoring algorithm based on three approaches for optimizing resource usage. The used approaches provide a deep knowledge about the person’s context by considering: the person’s profile, the activities and the relationships between activities. We evaluate the performances of our framework and show its adaptability and significant reduction in network, energy and processing usage over a traditional monitoring implementation.
\end{abstract}

\begin{IEEEkeywords}
Adaptive monitoring; context-aware monitoring; smart sensing; dependency; e-health; ADL; IADL
\end{IEEEkeywords}

%

\section{Introduction}

Health Smart Home (HSH) and Healthcare Information Systems (HIS) for elderly persons seek to monitor their illness, handicap, and dependence seamlessly and to provide e-health services that meet their context and real needs in their own homes. The main objectives of HSH are to maintain the level of dependency and delay recourse to establishments of healthcare (e.g. nursing homes, hospitals, etc.) as long as possible.
These objectives are achieved through a series of procedures and mechanisms including: environmental data acquisition (sensing and transmitting), data storage and analysis, making decision, then providing context-aware services. Context-aware e-health services refer to systems that can automatically acquire a person's information (environment, physical, activities), and are able to provide and adapt their services accordingly. Although a variety of sensor technologies are widely available today, context-aware e-health services systems do not satisfy the desired requirements and there are many aspects needing further improvement.

In order to be successful, context-aware e-health systems must have a full visibility of the person's context including what, when and how to monitor, gather and analyze data. In our previous work \cite{4}, we improved the knowledge about the context of the person which is the set of activities of daily living that should be monitored in such HIS, and show how they affect the performances of health monitoring systems. In this paper, we propose a framework for monitoring the different activities with an optimized use of resources (network, energy and processing) and without compromising the quality of the monitoring service and maintaining the system's ability to detect abnormal situations. The monitoring is dynamically adapted to the context and situation of the monitored person and his history, the nature of monitored activities and exiting relationships between activities.

\section{Related Work}

Most of the previous studies and projects in HSH/HIS operate in isolation from the real requirements of the healthcare institutions. Which in turn contributes to a high incidence of unsuccessful projects \cite{2}. The absence of such link leads to high uncertainty in the adoption of such projects. Therefore, it is desirable to improve the context-aware e-health systems and make easier the integration of the new proposed e-health systems into health institutions. The traditional health monitoring approaches tend to manage all sensed data with unconditional processing.  Most of these approaches are aimed at continuous monitoring while maintaining the channels of transfer data  available all the time. The adoption of such trend is causing many issues, such as collapse of the network, data transmission failure, energy consumption, important 
computational cost, and loss of priority in processing and making quick relevant decisions. The same situation happens with physical monitoring applications which assume uniform time-interval signal data.

Many systems and applications have been designed for dealing with different ideas including sensing and network communications \cite{5} \cite{6}, computing analysis algorithms \cite{8}, energy consumption \cite{12}, time-cost effectiveness, assisting and providing services \cite{13}, etc.  In existing HSH and HIS, there are three main categories: physiological monitoring, movement or fall detection and activity monitoring. Physiological systems record the vital signs of elderly and chronically ill \cite{14}. Movement or fall detection systems are ambulatory activities detection including dynamic activities (like $walking$) and static postures (like $standing$), location tracking and accidental falls \cite{15}.  Activity of Daily Living (ADL)
monitoring is covering a list of basic daily activities (such as $eating$ and $washing$) and allows the elderly to live independently and to provide accurate services for caregivers\cite{13} \cite{18}. Existing approaches lack a true understanding about the person's activities that should be considered in monitoring. Their lack of adaptability lead to a limited and narrow range of activity selection.

Many heterogeneous sensors and devices are deployed in HSH (e.g. cameras, sound detectors, lights, door sensors, etc.). This whole equipment increases both energy consumption and network traffic.
Saving power can be ensured by using scheduling for sensor nodes or sentry-based algorithms in order to maintain sensing coverage \cite{19}. However, such schemes are not automatically adjusted at the execution time of monitoring, therefore we need a dynamic choice of monitoring modes.
Some studies have resorted to specific types of wireless communication technology such as Zigbee \cite{18} or Bluetooth Low Energy (BLE) \cite{20} to reduce the energy consumption. This remains restrictive, since dealing with all available sensors and devices for full visibility of a person's context is imperative. Single low-power consumption sensors were used in \cite{21}, while \cite{22} used 167 sensors simultaneously. Therefore, we need to use efficient data-summarizing algorithms and relevant data filtering mechanisms, that consider all variables of a person's activities and environment. Dynamic updating of the monitoring mode, including moderate usage of sensors 
can help in detecting ADLs, which cannot be done with one unique sensor, but with a composition of several ones.

\section{Framework for e-health Monitoring}

\subsection{Data Processing Issues}

In the context of the diagnosis of health situations, the variation in the variables does not always mean a change in the diagnosis. In the context of e-health monitoring systems and elderly evaluation, the majority of the data processing and the results of the evaluation conclude that there is no relevant change since the previous evaluation.
Since the variability of the health condition is usually much lower than the variability of the signals by itself, health professionals usually perform sequence tasks interpretations based on data priority and limit the diagnosis set to the most relevant results \cite{23}.

Health monitoring which is performed either continuously or periodically with uniform time interval and unconditional processing, presents numerous problems. The continuous monitoring requires a huge amount of unnecessary sensed data to be sent to a coordinator (e.g. gateway) and processed without any conditions. For instance, continuously sensing the body weight. Consequently, the processing usually ends up with no change in the health status of the patient. One way to improve this approach is to define a new decision mechanism of monitoring based on the importance, priority and validity of the data (i.e. the period where a sensed value remains unchanged or where changes occur without any impact).
Moreover, periodical monitoring with fixed time interval leads to two unsatisfactory results. First, inequitable monitoring in severe dependency or critical health status, which needs the highest frequency of monitoring and data transmission. Second, wasted effort in cases of stable or non-critical situation.
Our target system aims to determine what are the activities and contextual dimensions (variables) that need to be monitored in an optimized way either for the continuous or periodic monitoring.

\subsection{Design}
\label{metho}

To reach our objectives, we propose a non-uniform interval that we call \emph{validity period} (VP). VP is an extension of the concept of priority data in the evaluation of needs. VPs are estimated on the basis of the person’s profile which includes: dependency situation, historical record, behavior, chronic diseases, etc. In order to update the validity of data and the frequency of data requests, we need different VPs: for each activity and within the same activity depending to the person’s health status and record. VPs are influenced by the detection of abnormal behavior. Such changes in patterns allow the HIS to notify caregivers in order to provide appropriate assistance and service. In order to achieve a robust hybrid approach able to solve previously mentioned problems, we propose an adaptive context-aware monitoring framework supported by a conditional processing scheme. The elderly profile which includes dependency level and historical record is the primary key for adapting the monitoring period of the sensor nodes, in order to collect and process highly relevant data.

\subsubsection{Framework Description}
\begin{figure}
\centerline{\includegraphics[totalheight=4cm]{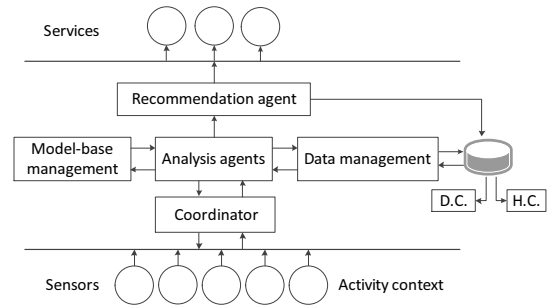}}
    \caption{Components of the Framework for e-health Monitoring}
    \label{Fig3. pic}
\end{figure}

the framework of the proposed approach is presented in Fig.~\ref{Fig3. pic}. The system uses the elderly activity context as a base for sensing, analyzing, and making service recommendation. The person's environment is equipped with a list of possible sensors which are placed in appropriate spaces. Data is obtained either continuously or periodically, then transmitted to the coordinator to be analyzed.
A data management system manages data coming from these several heterogeneous sources. It supports all the usual database primitives (e.g. add, delete, search, query, etc.).
The analyzing agent loads the person's profile including dependency-context (D.C.) and history-context (H.C.) from the data management system during a specified period.
 The analyzing agent connects to the model-base management, then the first inference is performed to start/configure sensors and set up the monitoring mode.  The model-base management is used to select the suitable models (e.g. SMAF \cite{25} and AGGIR \cite{13}) for the person's activities and his behavior, which in turn is provided to the analysis agent and data management to retrieve input data and set up outputs for monitoring.
Finally, after sensing and analyzing, the health services are recommended, based on the real needs of the monitored person.

\subsubsection{Approaches}
\label{Approaches}
in order to dynamically update the monitoring of the person’s activities, our adaptive context-aware monitoring system defines three approaches used together:
$Activity$ $per$ $Activity$ Approach, $Global$ Approach and $Relational$ Approach  (Sections \ref{catperact}, \ref{globalAp} and \ref{Relat}). Many existing models are used in the geriatrics domain to cover the human activities and provide a mean to evaluate the elderly ability in achieving them. The SMAF model is one of the most popular models that considers 29 based on the ADL and  \textit{Instrumental Activity of Daily Living} (IADL) concepts \cite{25}. We use the SMAF activities with additional ones in the definition of new monitoring modes. The monitoring modes consider additional information tailored to each activity. Such information is related to the monitoring frequency, duration, class of each activity, possible used sensors and scores (Table \ref{table1}). The frequency and duration are used to determine when and for how long the monitoring is performed. The classes of activity are defined to distinguish the consumption of network resources (e.g. traffic and power). Scores are used -as in any geriatric model- to quantify the ability of the person’s in achieving the activity.

\subsubsection{Activity per Activity Approach}
\label{catperact}
the objective of this approach is to consider each monitored activity separately. This approach takes into account the nature of each activity and the required time for the monitoring. This approach is important during the initial monitoring of autonomous persons and helps to learn the person’s behavior evolution. Table \ref{table1} presents a subset of the considered activities with some tailored information such as monitoring frequencies, durations and scores. As we can see, for each activity we set up an initial frequency ($x$ value) and duration for the monitoring of each activity separately. For instance, $x$ is fixed to 3 for the \textit{Toileting} activity (Table \ref{table1}), which means that we need to monitor this activity each 3 days. Consequently, we will have 10 monitoring results during a period $P$=30 days. This result will indicate how many times the person succeeded to achieve the activity correctly, let us call this value $activitySubScore$. We evaluate the levels of dependency using four modalities from score -3 to 0. So, we have -3: \textit{Dependence ($\mathcal{D}$)}, -2: \textit{Need help ($\mathcal{H}$)}, -1: \textit{Supervision ($\mathcal{S}$)} and 0: \textit{Autonomous ($\mathcal{A}$)}. Since we have 4 scores, the $activitySubScore$ value is tested within four intervals having a step of $P/4x$ where $P$ is the period chosen to re-evaluate the person’s dependency level. Example, with $P$ = 30 days and $x$ = 3 days, the score intervals are as follows: $\mathcal{D}$ $\equiv$ [0, $step$=2.5[, $\mathcal{H}$ $\equiv$ [$step$=2.5, $2.step$=5[, $\mathcal{S}$ $\equiv$ [$2.step$=5, $3.step$=7.5[, $\mathcal{A}$ $\equiv$ [$3.step$=7.5, $4.step$=10]. So, if the monitoring result is $activitySubScore$ = 8, the activity score ($activityScore$) will be $\mathcal{A}$ (i.e. autonomous).

The previous evaluation of the dependency level (i.e. $\mathcal{D}$, $\mathcal{H}$, $\mathcal{S}$ or $\mathcal{A}$) regarding each activity is the result related to a long period of monitoring $P$. The challenge is how to progressively (e.g. each 24 hours) compute $activitySubScore$ before reaching its final value. If we consider the previous example, the question is: how to obtain $activitySubScore$ = 8 for $P$ = 30 days? This requires to define a way where we judge if the person has performed well or not the activity during duration of monitoring. After this duration, if the activity is performed well, the system will count it (i.e. $C_{activity}$($duration$)=1), otherwise an abnormal situation is detected. For example, for some activities, we need to know how many times the activity was performed during a full day ($duration$ = 24 hours, Category II in Table \ref{table1}). For instance, for the \textit{Toileting} activity and according to the observed number, the system will decide if this activity was achieved correctly or not after the monitoring period as follows: if the observed number after 24 hours is 0 or 1 time, the situation is considered abnormal and the system will not count it i.e. the considered value will be zero ($C_{Toileting}$(24)=0). If the observed number is 2 or 3 times, the situation is judged normal and the considered value is $C_{Toileting}$(24)=1. Similarly, if the current number is higher than the previously observed average, the considered value is $C_{Toileting}$(24)=1. However, this last situation could be considered as abnormal depending to the person’s history (e.g. high \textit{toileting} activity with \textit{Diabetes} ; high recording of sleep disorders with \textit{Hepatitis C}, etc.). For any kind of abnormal detection, the system will extend the monitoring for an extra duration period to record opposite behavior and/or notify caregiver after receiving confirmation of the situation. For some activities, our monitoring system needs only to know if the activity is done or not in order to be counted. This is the case, for instance, of the \textit{Washing} activity (Category I in Table \ref{table1}) and other activities like \textit{housekeeping}, \textit{dressing} and \textit{laundry}.

\begin{table*}
\renewcommand{\arraystretch}{1.4}
  \caption{Activity per Activity Approach}
  \label{table1}
  \centering
  \begin{scriptsize}
  \begin{tabular}{*{7}{|c|@{}p{1.8cm}@{}|@{}p{2.6cm}@{}|@{}p{1.4cm}@{}|@{}p{1cm}@{}|@{}p{6.5cm}@{}|@{}p{3.8cm}@{}}}
   \hline
    &\textbf{ Activity} & \textbf{ Frequency}  & \textbf{ Duration} & \textbf{ Class} & \textbf{ Possible sensors or devices} & \textbf{ Scores} \\
    \hline
    \multirow{3}{*} {ADL}& \ Washing \ & \ each 10 days (i.e. $x$ = 10) \ & \ Category I$^1$ \ & \  medium \ &\ ultrasonic water flow meter, humidity, light, sound detector, temperature and/or motion sensor..\  & \  4 modalities$^4$ with a step of $P / (4x)$\ \\ \cline{2-1}\cline{3-1}\cline{4-1}\cline{5-1}\cline{6-1}\cline{7-1}
    & \ Toileting \ &\ each 3 days (i.e. $x$ = 3)\ &\  Category II$^2$ \ &\  medium \ &\  light, sound detector, flushes  switch, motion sensor and/or RFID.. \ &\ 4 modalities$^4$ with a step of $P / (4x)$ \ \\  
    \hline

   \multirow{1}{*}{Mob.}& \ Walking inside\  &\  N/A (computed) \ &\  Category III$^3$ \ &\  low  \ & \ floor plan, indoor GPS, motion sensor and/or RFID..\ & \  4 modalities$^4$ with a step of $P / (4x)$\  \\ 
    \hline

   \multirow{2}{*}{Com.}& \  Hearing \ &\  each 30 days ($x$=30)\  &\  Category III$^3$ \ &\ low  \ & \ TV questionaries’ achieved by the person \cite{13}\ & \  4 modalities$^4$\ \\ \cline{2-1}\cline{3-1}\cline{4-1}\cline{5-1}\cline{6-1}\cline{7-1}

   & \ Speaking \ & \ each 30 days ($x$=30)\  & \ Category III$^3$\  &\ low\  &\ TV questionaries’ achieved by the person..\ & \ 4 modalities$^4$ \ \\  
    \hline

      \multirow{2}{*}{M. F.}&\  Orientation  \ & \ N/A (computed) \ & \ \  & \  low \ &\  floor plan, indoor GPS, motion sensor and/or RFID..\ & \ 4 modalities$^4$ $Step = P /4$ \ \\ \cline{2-1}\cline{3-1}\cline{4-1}\cline{5-1}\cline{6-1}\cline{7-1}

      & \ Memory\   & \  each 30 days ($x$=30) \ & \ Category III$^3$  \ &\   low \ &\  TV questionaries’ achieved by the person.. \ &\   4 modalities$^4$\ \\ 
      \hline

        \multirow{2}{*}{IADL}&\  Housekeeping  \ &\  each 10 days($x$=10)\  &\ Category I$^1$ \  &\  medium \ & \ power sensor, sound detector and/or motion sensor..\ & \  4 modalities$^4$ with a step of $P / (4x) $ \ \\  \cline{2-1}\cline{3-1}\cline{4-1}\cline{5-1}\cline{6-1}\cline{7-1}
        & \ Meal preparation \  &\  each 3 days ($x$=3) \ & \ Category II$^2$ \ & \ high \ &\ washing dishes, mixer tap, gas, oven, toaster, light switch, door sensors and IP camera, sound \& motion detector, RFID..\  &\  4 modalities$^4$ with a step of $ P / (4x)$ \ \\ 
       \hline

        \multirow{2}{*}{Other}& \ Watching TV \  & \ each 10 days ($x$=10)\  &\ Category I$^1$ \  & \ high \ & \ power sensor, pressure sensor, IP camera and/or sound detector..\ &  \ 4 modalities$^4$ with a step of $P / (4x)$\ \\ \cline{2-1}\cline{3-1}\cline{4-1}\cline{5-1}\cline{6-1}\cline{7-1}
        & \  Sleeping \ &  \ each 3 days ($x$=3) \ &\  Category II$^2$ \ &\  low \ &\ pressure sensor and motion sensor..\  &\ 4 modalities$^4$ \  \\ 
       \hline

  \end{tabular}
  \end{scriptsize}
  \rule{0in}{1.2em}\scriptsize$^1$ sensor is active till the activity occurs, then the next monitoring will be after the $x$ period, \scriptsize$^2$ when the activity occurs, the monitoring will be during the next 24h , \scriptsize$^3$ the evaluation is based on the \emph{Relational Approach} (see Section \ref{Relat}), \scriptsize $^4$ possible scores are : 0, -1, -2 and -3\\
\end{table*}

\subsubsection{Global Approach}
\label{globalAp}
the objective of this approach is to avoid the exaggerations of the dependency evaluation models that requires the computation of all the activities at all the levels of dependency even in severe ones. Moreover, the system has to determine the degree of data frequency to avoid unnecessary data sensing, which is usually found in traditional monitoring systems as we studied it in our previous investigation \cite{4}. Request-driven monitoring approaches \cite{23} represent a good candidate to optimize the continuous monitoring. Therefore, our adaptive monitoring considers the global evaluation of the person as a base to increase or decrease the frequency of the monitoring. This approach is based on the initial $x$ value discussed previously and applying a global evaluation of the person abilities by using one of the existing geriatrics evaluation models. The idea is to provide a  monitoring with a dynamic frequency and activation depending to the level of dependency of the person (from autonomous to dependent) and the nature of the monitored activities (such as \textit{basic} activities –ADL and \textit{instrumental} activities –IADL). We ensure this dynamic monitoring by updating the $x$ value in order to expand or reduce the scope of the monitoring and sensed data.

For the present approach and based on our previous work \cite{4}, we adopt the SMAF evaluation model \cite{25}. 
Functional Autonomy Measurement System (SMAF) is a clinical rating scale that measures the functional autonomy of elderly patients. The SMAF used in order to rehabilitate the individual by provide appropriate care and services and assessing needs to alleviate the disabilities in elderly people. There are 29-items rating scale used to evaluates person’s dependency, and access to the available resources that may offset for the disabilities, as well the stability of resources. These items are included in the five aspects of functional abilities: activities of daily living ADL (7 items), mobility (6 items), communication (3 items), mental functions (5 items), and instrumental activities of daily living IADL (8 items).  The SMAF model is administrated manually by a health professional. The raters use all available information to do the rating. Dependency is evaluated by using a scale for each item with a 5-point rating scale: 0, -0.5, -1, -2 and -3. Items are evaluated using a function scoring: 0: independently, -0.5: independently but with difficulty, -1: needs supervision or stimulation,-2: needs help,-3: dependent.  The disability from autonomy to dependency is identified with a maximum negative scores of -87, a higher disability score indicates a higher level of dependence. The handicap assessment is necessary to overcome the disability score. If the social resources are accessible to compensate for the disability,  the handicap score is zero; otherwise the handicap score equals the disability score\cite{smaf2}.
SMAF has been developed in \cite{smaf3} to include 14 profiles of dependency patterns called iso-SMAF profiles. Each profile is associated with a specific amount of nursing, support services, supervision needed and the costs of services, based on the disabilities of their patient groups. In SMAF, the first profile ($P_{1}$) represents the persons that are autonomous while the last profile ($P_{14}$) represents completely dependent persons. These profiles based on the results of the information of all the 29 items. From the first to the last iso-SMAF profiles, the mean level of disability increases from 9.4 to 73.8 out of a potential of 87. These 14 iso-SMAF profiles divided into 4 categories: 1- includes subjects who are autonomous with some IADL required supervision and help ($P$ 1, 2, and 3); 2- includes subjects who show mobility functions disabilities ($P$ 4, 6, and 9); 3- includes subjects who show mental disabilities ($P$ 5, 7, 8, and 10); and 4- shows the lowest level of autonomy i.e. dependency in all ADL activities ($P$ 11, 12, 13, and 14) \cite{smaf3}\cite{25}. Table \ref{smaf} briefly illustrates the association between profiles, disability score and classification group of iso-SMAF profiles.

\begin{table}
\renewcommand{\arraystretch}{1.3}
\caption{Association between profiles, disability score and classification group in iso-smaf profiles}
\label{smaf}
\centering
\begin{scriptsize}
\begin{tabular}{|@{}p{1.1cm}@{}|@{}p{1.1cm}@{}|@{}p{1cm}@{}|@{}p{1.1cm}@{}|@{}p{1.1cm}@{}|@{}p{1cm}@{}|}
    \hline

Number of Profiles&Disability score /87& Category &  Number of Profiles& Disability score /87& Category\\ \hline

\ 1&\ 	-9.33&\ 	1&\ 	8&\ 	-42.24&\ 	3\\ \hline
\ 2&\ 	-13.23&\ 	1&\ 	9&\ 	-48.15&\ 	2\\ \hline
\ 3&\ 	-19.76&\ 	1&\ 	10&\ 	-53.02&\ 	3\\ \hline
\ 4&\ 	-23.69&\ 	2&\ 	11&\ 	-58.47&\ 	4\\ \hline
\ 5&\ 	-28.54&\ 	3&\ 	12&\ 	-58.71&\ 	4\\ \hline
\ 6&\ 	-32.04&\ 	2&\ 	13&\ 	-64.98&\ 	4\\ \hline
\ 7&\ 	-39.19&\ 	3&\ 	14&\    -73.77&\ 	4\\ \hline

\end{tabular}
\end{scriptsize}
\end{table}

Table \ref{table2} shows our system’s dynamic updates of the $x$ value regarding the current person’s profile, which is periodically evaluated. As we can see, the value of $x$ depends to the category of the activity (ADL, Mobility, etc.) and the modality of the dependency as defined in the SMAF model (autonomous, difficulties, etc.). For instance, for autonomous persons ($P_{1}$ and $P_{2}$), the system uses the default $x$ value (initialized in the previous approach) for the IADL activities and ignores the others (i.e. $x$ = $infinity$). Notice that for dependent persons, starting from $P_{6}$, all the activities are monitored with the default value of $x$ and with a high frequency of monitoring the IADL activities. This is explained by the fact that persons belonging to profile $P_{6}$ are suffering from a high dependency in achieving their IADL activities such as meal preparation and medication use.

Our general rule, within the same category of activities, is when the dependency increases, the $x$ value decreases except for the IADL activities in profiles $P_{7}$ to $P_{14}$ as we can see in Table \ref{table2}. Indeed, since IADL activities are lost first, this means that a person who is not able to achieve the IADL activities will start to need help in achieving the other kind of activities. This situation starts from $P_{7}$ (Table \ref{table2}). Consequently, we start decreasing the $x$ value for IADL from the profile $P_{7}$. For, and only for, the IADL activities, the $x$ value starts to become $infinity$ (i.e. no monitoring is needed anymore) when the person starts to need help in achieving the majority of the activities: from $P_{11}$ to $P_{14}$ where the person enters to the called \textit{long-term care facilities} (LTCF).

\begin{table}
\renewcommand{\arraystretch}{1.3}
\caption{Global Approach Monitoring}
\label{table2}
\centering
\begin{scriptsize}
\begin{tabular}{|@{}p{0.5cm}@{}|c|c|c|c|c|}
    \hline
          &  \textbf{ADL} & \textbf{Mob. }& \textbf{Com.} & \textbf{M.F.}& \textbf{IADL}\\ \hline
    \ P$_1$ & $x = inf.$ & $x = inf.$ & $x = inf.$ & $x = inf.$ & \cellcolor[gray]{.8}$x = x/1 $ (initial $x$)\\ \hline
    \ P$_2$ & $x = inf.$ & $x = inf.$ & $x = inf.$ & $x = inf.$ & \cellcolor[gray]{.8} $x = x/1$ \\ \hline
    \ P$_3$ & $x = inf.$ & $x = inf.$ & \cellcolor[gray]{.8} $x = x/1$ & \cellcolor[gray]{.8} $x = x/1$ & \cellcolor[gray]{.6}$x = x/2$ \\ \hline
    \ P$_4$ & $x = inf.$ & \cellcolor[gray]{.8} $x = x/1$ & \cellcolor[gray]{.8} $x = x/1$ & $x = inf.$ & \cellcolor[gray]{.6}$x = x/2$ \\ \hline
    \ P$_5$ & \cellcolor[gray]{.8} $x = x/1$ & $x = inf.$ & \cellcolor[gray]{.8} $x = x/1$ & \cellcolor[gray]{.8} $x = x/1$ &  \cellcolor[gray]{.1} {\color{white}$x = x/3$} \\ \hline
    \ P$_6$ & \cellcolor[gray]{.8} $x = x/1$ & \cellcolor[gray]{.8} $x = x/1$ & \cellcolor[gray]{.8} $x = x/1$ & \cellcolor[gray]{.8} $x = x/1$ & \cellcolor[gray]{.1} {\color{white}$x = x/3$}\\ \hline
    \ P$_7$ & \cellcolor[gray]{.8} $x = x/1$ & \cellcolor[gray]{.8} $x = x/1$ & \cellcolor[gray]{.8} $x = x/1$ & \cellcolor[gray]{.6}$x = x/2$ & \cellcolor[gray]{.1} {\color{white}$x = x/(3-1)$}\\ \hline
    \ P$_8$ & \cellcolor[gray]{.8} $x = x/1$ & \cellcolor[gray]{.8} $x = x/1$ & \cellcolor[gray]{.8} $x = x/1$ & \cellcolor[gray]{.6}$x = x/2$ & \cellcolor[gray]{.1} {\color{white}$x = x/(3-1)$}\\ \hline
    \ P$_9$ & \cellcolor[gray]{.6} $x = x/2$ & \cellcolor[gray]{.6} $x = x/2$ & \cellcolor[gray]{.8} $x = x/1$ & \cellcolor[gray]{.8}$x = x/1$ & \cellcolor[gray]{.1} {\color{white}$x = x/(3-2)$}\\ \hline
    \ P$_{10}$ & \cellcolor[gray]{.6} $x = x/2$ & \cellcolor[gray]{.8} $x = x/1$ & \cellcolor[gray]{.8} $x = x/1$ & \cellcolor[gray]{.6}$x = x/2$ & \cellcolor[gray]{.1} {\color{white}$x = x/(3-2)$}\\ \hline
    \ P$_{11}$ & \cellcolor[gray]{.6} $x = x /2$ & \cellcolor[gray]{.6} $x = x/2$ & \cellcolor[gray]{.8} $x = x/1$ & \cellcolor[gray]{.6}$x = x/2$ & \cellcolor[gray]{.1} {\color{white}$x = inf.$}\\ \hline
    \ P$_{12}$ & \cellcolor[gray]{.1} {\color{white}$x = x/3$} & \cellcolor[gray]{.6} $x = x/2$ & \cellcolor[gray]{.8} $x = x/1$ & \cellcolor[gray]{.6}$x = x/2$ & \cellcolor[gray]{.1} {\color{white}$x = inf.$}\\ \hline
    \ P$_{13}$ & \cellcolor[gray]{.1} {\color{white}$x = x/3$} & \cellcolor[gray]{.1} {\color{white}$x = x/3$} & \cellcolor[gray]{.8} $x = x/1$ & \cellcolor[gray]{.6}$x = x/2$ & \cellcolor[gray]{.1} {\color{white}$x = inf.$}\\ \hline
    \ P$_{14}$ & \cellcolor[gray]{.1} {\color{white}$x = x/3$} & \cellcolor[gray]{.1} {\color{white}$x = x/3$} & \cellcolor[gray]{.6} $x = x/2$ & \cellcolor[gray]{.1} {\color{white}$x = x/3$} & \cellcolor[gray]{.1} {\color{white}$x = inf.$}\\ \hline

\end{tabular}
\end{scriptsize}
\fboxsep=.5mm \fboxrule=.1mm
\fcolorbox{black}{white!40}  Autonomy,
\fcolorbox{black}{white!40}  Difficulties,
\fcolorbox{black}{black!20}  Supervision,
\fcolorbox{black}{black!40}  Help,
\fcolorbox{black}{black!80} {\color{white}D}ependence
\rule{0in}{1.2em}\scriptsize \cite{25}\\
\end{table}

\subsubsection{Relational Approach}
\label{Relat}
\begin{figure}
\centerline{\includegraphics[totalheight=2cm]{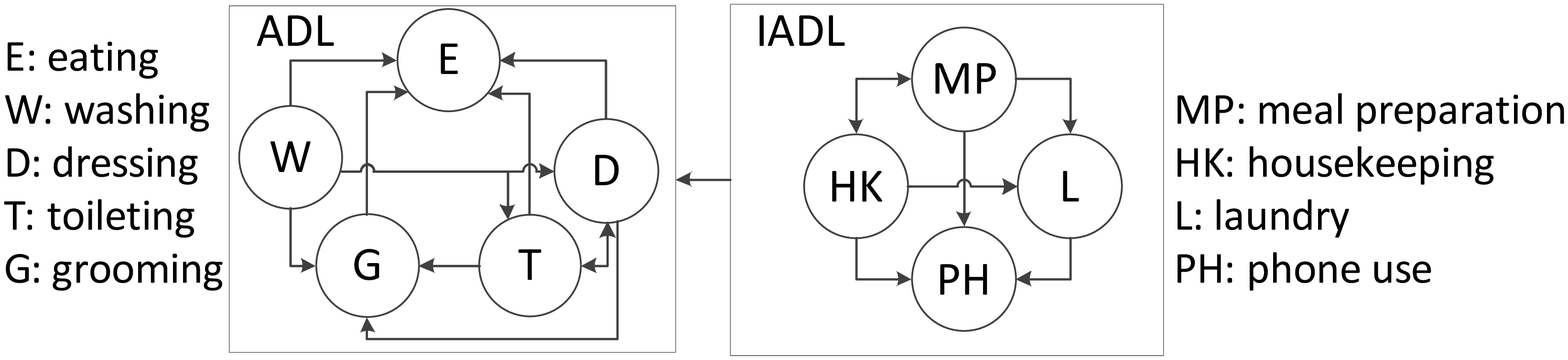}}
    \caption[figure4]{Relations between activities.}
    \label{figure4}
\end{figure}

this approach takes advantage of the two previous ones and focuses on the logical relationships that can exist between the monitored activities. Based on the existing logic related to the person's ability in performing different activities, we can improve the monitoring mode and optimize the system resources without compromising the level and quality of the monitoring service and maintaining the system’s ability to detect abnormal situations. The digraph $\mathcal{G}$ = ($V$, $A$), presented in Fig. \ref{figure4}., identifies a subset of the identified relationships between functional abilities, $a_{i}$ $\in$ $V$, within and between the ADL and IADL categories.
An arc $e$ = ($a_{i}$, $a_{j}$) means that if the person is able to achieve correctly the $a_{i}$ activity so he is able to achieve correctly the $a_{j}$ activity. Consequently, if the system detects a correct ability in achieving the activity $a_{i}$, no monitoring will be needed regarding the $a_{j}$ activity hence the $x$ value of $a_{j}$ is set to $infinity$ till the next detection of any change regarding the person’s ability.

\subsection{Scenario Generation}

In order to validate our previous approaches, there is a need of input scenarios that describe the activities of daily living of a given single person. One of the identified requirements is that the scenario should provide activities and events in a realistic manner: i.e. close to a real life of an elderly. Another requirement is to have a scenario for a long time period, e.g. a whole year, in order to be able to test the different facets of our approaches such as the dynamic adaptability of the monitoring. Real-life architectures imply complex implementations that include an important number of sensors required for considering the long list of activities (e.g. those defined by geriatrics models). A real-life implementation requires a long-time testing of the proposed approaches within an elderly using a fixed set of sensors. This represents a lack of flexibility to vary the used sensors and may represents a risk for the resident if the algorithms should evolve during the testing. A final requirement for the input scenario is to have some abnormal situations that should occur during the monitoring period.

To fulfill the previous requirements, it becomes clear that for an efficient evaluation, we need a rich, realistic and flexible scenario that considers different person activities and for a long period. Unfortunately, most existing works for scenario datasets did not consider a rich set of activities and fail to provide a good visibility regarding the person's context \cite{27}. For this work, we define a new scenario generation strategy based on Markovian models. The objective is to generate a long and rich sequence of activities for an elderly with or without disabilities. Specifically, we use the class of variable-length Markov models (VMM \cite{VMM}) in order to gain a certain expressivity during the generation of sequences $s = a_{1}, a_{2}, . . ., a_{l}$  of daily living activities with an order $l$ greater than one. The set of $a_{i}$ represent human actions required to perform the different activities as defined in geriatric models, mainly using the SMAF model \cite{25}. To an action $a_{i}$ a $startingTime$ and a random duration $activityDuration$ $\in$ [$aD_{min}$, $aD_{max}$] are associated. Also, we define a random $transitionTime\ (a_{i}, a_{j})$ $\in$ [$tT_{min}$, $tT_{max}$] from the end of an action $a_{i}$ to the starting time of an action $a_{j}$. In order to simplify our generation process while outputting realistic sequences, we define 5 transition matrices associated to the following periods of a day: 8.00 am - 11.00 am, 11.00 am - 2.00 pm, 2.00 pm - 5.00 pm, 5.00 pm - 10.00 pm and 10.00 pm - 8.00 am. Each matrix gives a higher probability to the most activities that can be met during a given period such as \textit{taking a shower} in the morning and \textit{eating} between 11.00am and 2.00 pm. Moreover, we define two additional transition matrices dedicated to Friday and Sunday since these days could include some specific activities with higher probabilities such a possible \textit{housekeeping} in Friday and a possible \textit{go outside} for a long period in Sunday. Our matrices and generated scenarios can be accessed online on \cite{ourDataProject}.

The generation strategy follows the Markov property by exploring a transition matrix $\mathcal{M}_{p\in\{1\ to\ 7\}}$, where a state represents a possible action $a_{i}$, and generating a future activity $a_{j}$ depending to the probability of $P$\ ($a_{j}$ $\mid$ $a_{i}$) = $\mathcal{M}_{p}$\ ($a_{i}$, $a_{j}$). Each generated $a_{j}$ is appended to the current sequence $s$. This \textit{random walk} suffers from some drawbacks. Mainly, the possible generation of less probable sequences and the lack of control during the sequences' generation. For instance, a \textit{blind} random walk approach could generate a sequence without a necessary activity in a given period of a day or a sequence that takes a long time that significantly exceeds a given period. To overcome these concerns, we control our generation strategy by introducing a set of constraints, which leads us to follow a \textit{pseudo} Markovian model where the sequence generation follows the transition probabilities under certain conditions. Constraints are checked for every possible transition from $a_{i}$ to $a_{j}$. If adding $a_{j}$ to the current sequence violates the constraints, the controlled random walk generates another possible activity $a_{j'}$. By construction, $\mathcal{M}_{p}$ matrices and added constraints guarantee the following properties: (a) the generation of finite and convergent sequences and (b) the use of transitions that are faithful to the dependency level of the person based on the SMAF definition of profiles \cite{25}. Two main constraints are used: the frequency $f$($a_{i}$) $\in$ [$f_{min}$, $f_{max}$] of some actions and the total sequence duration $sd$. To respect the frequency constraints, the scenario generation should ensure that $a_{i}$ appears at least $f_{min}$ time (that could be null) and do not exceed $f_{max}$. This is used to control how some actions should appear in $s$ (e.g. the number of \textit{eating} and \textit{washing} for an autonomous person). Once all the non-zero $f(a_{i})_{min}$ are satisfied, the generation stops when the total sequence duration reaches or starts to exceed the value of $sd$ = $\sum_{i=1}^{l} activityDuration\ (a_{i}) + \sum_{i=1}^{l-1} transitionTime\ (a_{i},a_{i+1})$ for $s = a_{1}, a_{2}, . . ., a_{l}$.

\subsection{Proposed Algorithm}

We summarize in Fig.~\ref{alg1} our adaptive and context-aware algorithm for monitoring the activities of daily living of elderly and dependent persons. The idea is to use the different approaches, discussed previously, in order to provide a monitoring that is dynamically adapted to the current situation of the person. The algorithm simulates the time evolution and applies the previous approaches on different input scenarios generated for one year \cite{ourDataProject}. We consider 29 activities (Table \ref{table1}) with 9 sub activities such as \emph{washing hand/face}, \emph{hair dry} and \emph{makeup} for the \emph{grooming} activity ; \emph{wash dish}, {make coffee}, \emph{make tea}, \emph{make sandwich} (toaster), \emph{make hot food} (microwave) and \emph{move dish} for the \emph{meal preparation} activity. The $x$ value, related to the frequency of the monitoring, depends on the nature of the monitored activity and is updated regarding to the evolution of the context (profile) of the person, such as the loss of abilities in achieving some tasks. These abilities are computed using scores associated to the different activities (see Algorithm 1, lines 9, 16 and 29) and a global score related to the profile of the person (line 32). Our second and third approaches are implemented using the $GlobalApproachUpdates$ and $RelationalApproachUpdates$ functions shown in Algorithm 2. Our algorithm uses adaptive monitoring periods with the required duration and includes the determination of the next monitoring time and the frequency in which the used sensors should send their data. The algorithm allows the evaluation of the network traffic and energy consumption implied by the adaptive monitoring (lines 10 and 17).

\alglanguage{pseudocode}
\begin{algorithm}
\caption{Adaptive Monitoring}
\label{alg1}
\small
\begin{algorithmic}[1]
\Procedure{$\mathbf{AdaptiveMonitoring}$}{}
    \State $A \gets 29\ activities$; $N \gets 365*24*3600\ seconds;$
    \State $activity\gets\emph{readLine(inputScenario)};$ $nxtMTime(a_{i}) \gets 0;$
    \LineComment{\emph{{\footnotesize read the first activity \& initialize "next monitoring time" for activities}}}

    \For {$i = 1 \to N$}
    \LineComment{\emph{{\footnotesize simulate the time evolution, $i$ is the current instant}}}

            \If {$i$ == $startingTime(activity)$}
			   \Switch{$activity$}
			   \LineComment{\emph{{\footnotesize see Section \ref{Approaches} \& Table \ref{table1} for Categories}}}
			      \Case{Category I} \textbf{: }
			                   \If {$i \geq  nxtMTime (activity)$}
			                   		 \State $activitySubScore(activity)${\footnotesize ++};
			                   		\State \emph{compute network traffic and power consumption};
			                   		\State \emph{updates $nxtMTime (activity)$};
			                   \EndIf
			     \EndCase
			     \Case{Category II} \textbf{: }
			                   \If {$i \geq  nxtMTime (activity)$ \textbf{and} \\
			                   \ \ \ \ \ \ \ \ \ \ \ \ \ \ \ \ \ \ \ \ \ \ \ \ $i \leq  nxtMTime (activity) + 24h$}
			                   		 \State $temporaryActivitySubScore(activity)${\footnotesize ++};
			                   		 \LineComment{\emph{{\footnotesize after 24h activitySubScore will be computed}}}
			                   		\State \emph{compute network traffic and power consumption};

			                   \EndIf
			     \EndCase
			   \EndSwitch
			   \State $activity \gets \emph{readLine (inputScenario)};$
            \EndIf
            \For{\textbf{each} $a$ in Category II}
            	\If {$i \geq  nxtMTime (a)+24h$}
            		\State \emph{{\footnotesize computeActivitySubScore (a)}};
            		\LineComment{\emph{{\footnotesize after 24h activitySubScore is computed}}}
            		\LineComment{\emph{{\footnotesize using $temporaryActivitySubScore(a)$}}}
            		\State \emph{updates $nxtMTime (a)$};
            	\EndIf
            \EndFor
             \If {mod ($i$, 30 days) == 0}
                         		 \LineComment{\emph{{\footnotesize each month, the activityScore is computed (see Section \ref{Approaches})}}}
             		 \For {$l = 1 \to A$}
             			\State $activityScore(a_{l}) \gets $\/
             			\State $SMAFScore(activitySubScore(a_{l}));$
             		
             		 \EndFor
             		\State $profile \gets computeSMAFProfile(activityScores)$;
             		\LineComment{\emph{{\footnotesize for computing the SMAF score, see \cite{4} }}}
             		\State $GlobalApproachUpdates(profile);$
             		\State $RelationalApproachUpdates(activityScores);$
             \EndIf

    \EndFor

\EndProcedure
\Statex
\end{algorithmic}
\vspace{-0.4cm}
\end{algorithm}

\alglanguage{pseudocode}
\begin{algorithm}
\caption{Helper Functions and Procedures}
\label{alg2}
\small
\begin{algorithmic}[1]
\Function{$\mathbf{SMAFScore}$}{$activitySubScore(a_{l})$}
	\State $P \gets 30\ days;$ $step \gets P/4\mathcal{X} \_Value(a_{l});$
	
	\LineComment{\emph{{\footnotesize see Section \ref{catperact}}}}
	\State $v \gets activitySubScore(a_{l});$
	\Switch{$v$}
		\Case{$v \geq 0$ and $v < step$} \textbf{: return} -3;
	    \EndCase
	    \Case{$v \geq step$ and $v < 2.step$} \textbf{: return} -2;
	    \EndCase\\
	   \ \ \ \ \ \ \ \ \ . . .
    \EndSwitch		
\EndFunction
\end{algorithmic}
\begin{algorithmic}[1]
\Procedure{$\mathbf{GlobalApproachUpdates}$}{$profile$}
	\Switch{$profile$}
		\Case{$P_{1}$} \textbf{: }
               \State updates $nxtMTime(a_{i})$; updates $\mathcal{X} \_Value(a_{i})$;
	    		\LineComment{\emph{{\footnotesize this is applied for all the activities $a_{i}$ (Table II)}}}
	    \EndCase
	    \Case{$P_{2}$} \textbf{: } . . .
	    \EndCase
    \EndSwitch	
\EndProcedure
\end{algorithmic}
\begin{algorithmic}[1]
\Procedure{$\mathbf{RelationalApproachUpdates}$}{$activityScores$}
	\If {$activityScores(a_{i})$ == 0} updates $nxtMTime(a_{j})$;
		\LineComment{\emph{{\footnotesize If $a_{i}$ is achieved autonomously, do not}}}
		\LineComment{\emph{{\footnotesize monitor related activities $a_{j}$, see Section \ref{Relat}}}}
	\EndIf
\EndProcedure
\Statex
\end{algorithmic}
\vspace{-0.4cm}
\end{algorithm}

\section{Experimentation}

We have conducted a number of simulations for the outcome of the person's behavior for a whole year. Notice that our proposed framework is not limited to the monitoring of a single person. Indeed, once the actor of a given activity is identified the proposed approaches remain the same. The person’s identification can be guaranteed using any kind of techniques such as RFID. However, for the sake of simplicity and to focus on the adaptive monitoring issues, we perform our experimentations based on the monitoring of one single person.

We apply the previous algorithm, with the discussed three approaches, in different scenarios within the same person's profile ($P_{1}$) and with simulating the person’s loss of abilities (with profile changes). Performed experimentations with profile changes focus on the adaptability of our approach and compare the monitoring cost to the situation where the person's level of dependency is stable. We evaluate the optimization of our adaptive monitoring that reduces sensing data without compromising to the credibility and reliability of dependency evaluation and with the identification of abnormal situations that may cause a risk for the person. Theses evaluations measure: the computing process (number of monitored activities), the detection of abnormal situations, the energy and network traffic consumption for a traditional (continuous) monitoring and our adaptive monitoring. In order to avoid a restrictive evaluation specifically for network traffic and energy consumption, we consider three classes of sensor nodes: $high$, $medium$ and $low$ used in monitoring the person’s activities (Section \ref{Approaches}). Resources consumption depends on the nature of the sensor used to monitor a given activity (Table \ref{table1}). For instance, for the \emph{low} class, we consider typical sensors with standard power values: 10.8mA, 7.5mA and 1$\mu$A in the transmitting, idle/receiving and sleeping modes respectively \cite{refSensor}.

\begin{figure}
\centerline{\includegraphics[width=8cm]{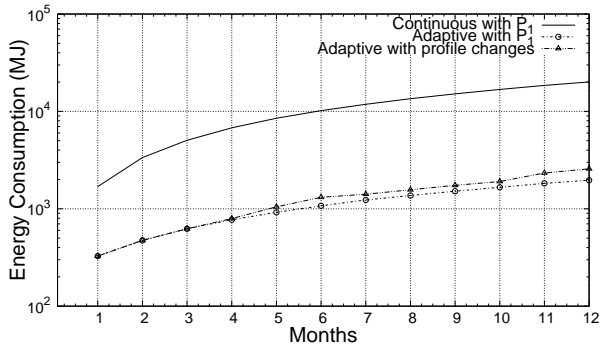}}
    \caption{Accumulated Energy Consumption}
    \label{fig1}
\end{figure}

\begin{figure}
\centerline{\includegraphics[width=8cm]{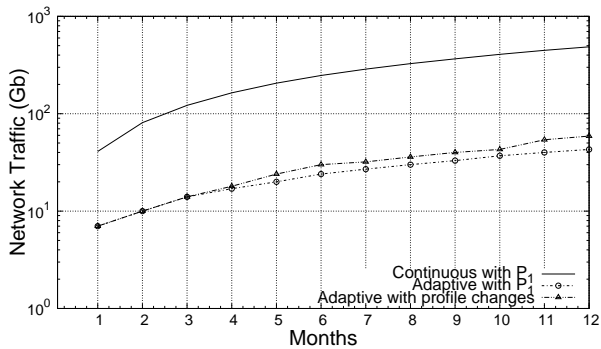}}
    \caption{Accumulated Network Bandwidth Consumption}
    \label{fig2}
\end{figure}

Figures \ref{fig1} and \ref{fig2} respectively compare the accumulated energy and network traffic consumption between a continuous and our adaptive monitoring within the same person’s profile ($P_{1}$). The observed gain (90.2\% for the energy and 90.99\% for the network traffic) is due to the conditional monitoring that considers the nature of the activities and the person's profile; hence no extra monitoring and data collection are used when it is not required. In order to evaluate the adaptability of our approach, we also compare the cost of our adaptive approach in two scenarios: within the same level of dependency (i.e. same profile) and with profile changes (see plots labeled \textit{Adaptive with $P_{1}$ and with profile changes} in Fig. \ref{fig1} and \ref{fig2}). Profile changes are: $P_{1}$ during the first 3 months, $P_{3}$ from the beginning of month 4 to month 6, $P_{6}$ from the beginning of month 7 to month 9 and profiles $P_{9}$ then $P_{8}$ then $P_{9}$ for the last three months.

\begin{figure}
\centering
\begin{tikzpicture}[y=0.14cm, x=.50cm,font=\sffamily]
\centering
	\draw (0,0) -- coordinate (x axis mid) (12,0);
    	\draw (0,0) -- coordinate (y axis mid) (0,18);
    	\foreach \x in {0,1,...,12}
     		\draw [very thin] (\x,.5pt) -- (\x,-.5pt)
			    node[anchor=north] {\tiny \x};
    	\foreach \y in {0,2,...,18}
         		\draw [very thin,dotted] (172pt,\y) -- (-.5pt,\y)
     			node[anchor=east] {\tiny \y};
	\node[below=0.2cm] at (x axis mid) {{\fontsize{0.25cm}{1em}\selectfont Months}};
	\node[rotate=90, above=0.3cm] at (y axis mid){{\fontsize{0.25cm}{1em}\selectfont \# of False Alarms}};
	\draw [thick] plot[mark=triangle*,  mark size=1.5pt, mark options={fill=white}]
		        file {mathematica.data5};
   \draw [thick] plot[mark=*,   mark size=1pt,  mark options={fill=white}]
		      file {mathematica.data6};
	\begin{scope}[shift={(0.1,13.5)}] 
	    \draw[yshift=\baselineskip] (0,0) --	
              plot[mark=triangle*, mark size=1pt, mark options={fill=white}] (0.25,0) -- (0.5,0)
               node[right]{\tiny  Continuous};
        \draw (0,0) --
	       plot[mark=*, mark size=1pt, mark options={fill=white}] (0.25,0) -- (0.5,0)
	       node[right]{\tiny Adaptive};

	
	\end{scope}
\end{tikzpicture}
\caption{Number of False Alarms}
\label{fig3}
\end{figure}
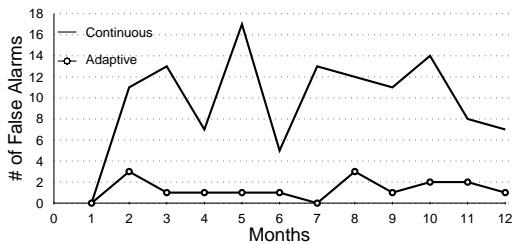

One of the common problems in continuous monitoring systems are the repetition of false alarms due to the person’s behavior changes. Example, for an \emph{autonomous} person, achieving a number of the same activity (like \emph{Toileting}) -that is significantly higher or smaller than usual- can be mistakenly considered as a risky situation. Our approach reduces the scope of this phenomenon (Fig. \ref{fig3}) since the person’s profile (i.e. the same health conditions) is considered. Hence, high numbers of useless alarms are avoided.

\begin{figure}
\centerline{\includegraphics[width=8cm]{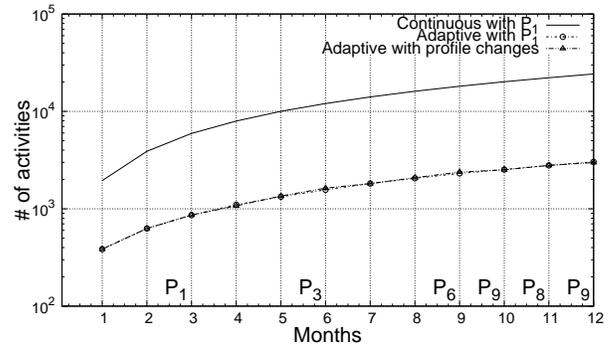}}
    \caption{Accumulated Number of Monitored Activities with Profile Changes}
\label{fig4}
\end{figure}

Elderly persons’ needs of assistance and services are changing gradually over the time, therefore the adaptation with these changes of person's life over long-term are required in e-health systems. However, this is not the only concern, indeed, it is of paramount importance to ensure a quick adaptation to sharp the sudden decline of the health status. Figure \ref{fig4} shows an increase of the monitoring of our system with three declines of the  health conditions: $P_{1}$ $\rightarrow$ $P_{3}$, $P_{3}$ $\rightarrow$ $P_{6}$ and $P_{6}$ $\rightarrow$ \{$P_{8}$, $P_{9}$\}. We observe a high accuracy of decline detection thanks to the consideration of the dependency level, the history of the behavior and the detection of abnormal situations while keeping a very low amount of sensed data. Notice that in spite of increasing the monitoring frequency when the dependency level increases, the observed number of monitored activities by our approach is almost the same with or without profile changes (see plots labeled \textit{Adaptive with $P_{1}$ and with profile changes} in Fig. \ref{fig4}). This is simply explained by the fact that the person achieves a reduced number of activities when he starts loosing its abilities.

\begin{figure}
\centering
\begin{tikzpicture}[y=0.045cm, x=.50cm,font=\sffamily]
\centering
	\draw (0,0) -- coordinate (x axis mid) (12,0);
    	\draw (0,0) -- coordinate (y axis mid) (0,70);
    	\foreach \x in {0,1,...,12}
     		\draw [very thin] (\x,.5pt) -- (\x,-.5pt)
			    node[anchor=north] {\tiny \x}; 
    	\foreach \y in {0,10,...,70}
         		\draw [very thin] (172pt,\y) -- (-.5pt,\y)
     			node[anchor=east] {\tiny \y};
	\node[below=0.2cm] at (x axis mid) {{\fontsize{0.25cm}{1em}\selectfont Months}};
	\node[rotate=90, above=0.3cm] at (y axis mid) {{\fontsize{0.25cm}{1em}\selectfont \# of abnormal detections}}; 
	\draw [thick] plot[mark=triangle*,  mark size=1.5pt, mark options={fill=white}]
	       file {mathematica.data9};
   \draw [thick] plot[mark=*,   mark size=1pt,  mark options={fill=white}]
             file {mathematica.data10};
	\begin{scope}[shift={(0.1,55.5)}] 
	   \draw[yshift=\baselineskip] (0,0) --	
             plot[mark=triangle*, mark size=1pt, mark options={fill=white}] (0.25,0) -- (0.5,0)
            node[right]{{\fontsize{0.17cm}{0cm}\selectfont Continuous}};
        \draw (0,0) --
	       plot[mark=*, mark size=1pt, mark options={fill=white}] (0.25,0) -- (0.5,0)
	       node[right]{{\fontsize{0.17cm}{0cm}\selectfont Adaptive} };

	

   \draw [dashed] (2.9,-56.9) -- (2.9, 18.5); 
   \draw [->] (-0.12, 17.0) -- (2.9, 17.0); 
   \draw (2.5, 19.4) node{{\fontsize{0.06cm}{1em}\selectfont $P_{1}$}};

   \draw [dashed] (5.9,-56.9) -- (5.9, 18.5); 
   \draw [->] (3.05, 17.0) -- (5.9, 17.0); 
   \draw (5.5, 19.4) node{{\fontsize{0.06cm}{1em}\selectfont $P_{3}$}};

   \draw [dashed] (8.9,-56.9) -- (8.9, 18.5); 
   \draw [->] (6.05, 17.0) -- (8.9, 17.0); 
    \draw (8.5, 19.4) node{{\fontsize{0.06cm}{1em}\selectfont $P_{6}$}};

   \draw [dashed] (9.9,-56.9) -- (9.9, 18.5); 
   \draw [->] (9.05, 17.0) -- (9.9, 17.0); 
    \draw (9.5, 19.4) node{{\fontsize{0.06cm}{1em}\selectfont $P_{9}$}};

   \draw [dashed] (10.9,-56.9) -- (10.9, 18.5); 
   \draw [->] (10.05, 17.0) -- (10.9, 17.0); 
    \draw (10.5, 19.4) node{{\fontsize{0.06cm}{1em}\selectfont $P_{8}$}};

   \draw [dashed] (11.9,-56.9) -- (11.9, 18.5); 
   \draw [->] (11.05, 17.0) -- (11.9, 17.0); 
    \draw (11.5, 19.4) node{{\fontsize{0.06cm}{1em}\selectfont $P_{9}$}};

	\end{scope}
\end{tikzpicture}
\caption{Detection of Abnormal Situations with Profile Changes}
\label{fig5}
\end{figure}
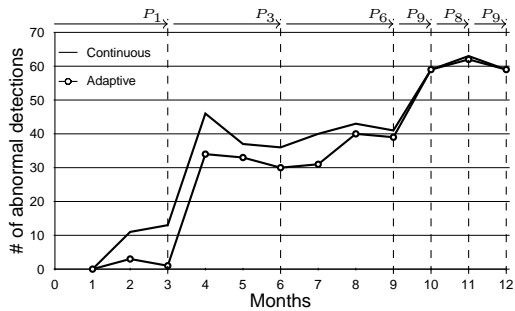

The results reveal that, even with declining elderly profiles, ensuring a timely and context-aware monitoring does not require sensing a huge amount of data. Indeed, our adaptive approach needs sensing only 15.12\% of data compared to a traditional continuous monitoring (Fig. \ref{fig4}). Hence an important saving of energy and network traffic. Table \ref{profilechange} shows the percentages of saving (sensing activities, power and network traffic) between traditional continuous monitoring and our adaptive system, we saved 84.9\% of sensing activities, 89.3\% of energy and 89.9\%  of network traffic. With this significant optimization of resources, the proposed approach succeeds to ensure a high accuracy (87.28\%) for the detection of abnormal situations (Fig. \ref{fig5}). Abnormal situations are generated in the input scenario by associating the activities with particular values regarding: durations (e.g. a very long or very short value for specific activities such as $toileting$) and frequencies (e.g. a null value for some required activities such as $eating$ and $medication$ $use$). Notice that the accuracy of 87.28\% represents an average and includes the period where the person is almost autonomous (from the first to the third month) where abnormal detection includes false alarms. The accuracy in detecting abnormal situation tends towards 100\% when there is a serious loss of the person's abilities (especially starting from month 8, Fig. \ref{fig5}) which confirms the efficient adaptability of our approach.

\begin{table}
\renewcommand{\arraystretch}{1.3}
\caption{Comparison between continuous and adaptive system (per activities and with profile changes)}
\label{profilechange}
\centering
\begin{scriptsize}
\begin{tabular}{|@{}p{2cm}@{}|@{}p{2cm}@{}|@{}p{1.8cm}@{}|@{}p{1.8cm}@{}|}
    \hline
& \ Sensing activities saving (\%) & \ Power saving (\%) & \ Network traffic saving (\%) \\ \hline

Eating & \ 93.9709	& \ 93.5675	& \ 93.5675 \\ \hline
Dressing& \ 98.8984	& \ 98.8179	& \ 98.8179 \\ \hline
Washing & \ 96.6292	& \ 96.9211	& \ 96.9211 \\ \hline
Grooming & \ 99.6508	& \ 99.6711	& \ 99.6711 \\ \hline
Toileting & \ 92.6705	& \ 92.6022	& \ 92.6022 \\ \hline
Housekeeping & \ 30.7692	& \ 32.3409	& \ 32.3409 \\ \hline
Laundry & \ 76.9231	& \ 76.4868	& \ 76.4868 \\ \hline
Meal preparation & \ 48.7549	& \ 54.9485	& \ 54.9485 \\ \hline
Telephone use & \ 95.7983	& \ N/A	& \ N/A \\ \hline
Medication use & \ 65.2174	& \ 65.4096 & \ 65.2174 \\ \hline
Watching TV & \ 96.8077	& \ 96.6794	& \ 96.6794 \\ \hline
Reading & \ 88.8601	& \ 86.1941	& \ 86.1941 \\ \hline
Sleeping & \ 66.0952	& \ 61.3398	& \ 63.2381 \\ \hline
Weight & \ 98.8571	& \ 98.8571	& \ 98.8571 \\ \hline
Walking inside & \ 81.8147	& \ N/A (computed)	& \ N/A (computed) \\ \hline
Total & \ 84.8809	& \ 89.3315	& \ 89.8829 \\ \hline

\end{tabular}
\end{scriptsize}
\end{table}

\section{Conclusion}

We have proposed in this paper, a context-aware adaptive framework for monitoring the activities of elderly daily living. The framework provides an efficient monitoring that optimizes the system resources. This efficiency is obtained by using conditional monitoring and processing, considering the person's profile to dynamically determine the frequency of data sensing, automatically evaluating the dependency, using abnormal behavior detection and power saving. Our experimentation, covering one year of data, has shown that our adaptive context-aware monitoring system determines the context of the person with a high accuracy (87,28\% of accuracy and more for severe dependency situations) and is able to sense data with a high flexibility while reducing the energy consumption (by 90.99\%) and the network bandwidth use (by 90.2\%).


\bibliographystyle{IEEEtran}
\bibliography{mshali-globecom2015}
\end{document}